\documentstyle[galley,grl,epsfig,amsmath]{agu2001}
\begin{document}
\title{Scaling of solar wind $\epsilon$ and the AU, AL and AE indices as seen
by WIND.}
\author{B. Hnat, \altaffilmark{1} S.C. Chapman \altaffilmark{1},
G. Rowlands \altaffilmark{1}, N.W. Watkins \altaffilmark{2} and
M.P. Freeman \altaffilmark{2}}
\altaffiltext{1}{Space and Astrophysics Group, University of Warwick Coventry, CV4 7AJ, UK}
\altaffiltext{2}{British Antarctic Survey, Natural Environment Research Council, Cambridge, CB3 0ET, UK}

\begin{abstract}
We apply the finite size scaling technique to quantify the statistical
properties of fluctuations in AU, AL and AE indices and in the $\epsilon$
parameter that represents energy input from the solar wind into the
magnetosphere. We find that the exponents needed to rescale the probability
density functions (PDF) of the fluctuations are the same to within
experimental error for all four quantities. This self-similarity persists
for time scales up to $\sim 4$ hours for AU, AL and $\epsilon$ and up to
$\sim 2$ hours for AE.
Fluctuations on shorter time scales than these are found to have similar
long-tailed (leptokurtic) PDF, consistent with an underlying turbulent
process. These quantitative and model-independent results place important
constraints on models for the coupled solar wind-magnetosphere system.
\end{abstract}

\begin{article}
\section{Introduction}
Recently, there has been considerable interest in viewing the coupled solar
wind-magnetosphere as a complex system where multi-scale coupling is a
fundamental aspect of the dynamics (see \citep{chang92b,chapman01} and
references therein). Examples of the observational motivation for this
approach are i) bursty transport events in the magnetotail \citep{angelopoulos}
and ii) evidence that the statistics of these events are self-similar (as seen
in auroral images \citep{lui}). Geomagnetic indices are of particular interest
in this context as they provide a global measure of magnetospheric output
and are evenly sampled over a long time interval. There is a wealth of
literature on the magnetosphere as an input-output system (see for example,
\citep{klimas,sitnov,tsurutani,vassiliadis,voros}).
Recent work has focussed on comparing some aspects of the scaling properties
of input parameters such as $\epsilon$ \citep{akasofu} and the AE index
\citep{davis} to establish whether, to the lowest order, they are directly
related \citep{freeman,uritsky}.
Although these studies are directed at understanding the coupled solar
wind-magnetosphere in the context of Self-Organized Criticality (SOC),
a comprehensive comparison of the scaling properties of the indices, and
some proxy for the driver ($\epsilon$) also has relevance for the
predictability of this magnetospheric ``output" from the input.
Importantly, both ``burstiness" (or intermittency) and self-similarity can
arise from several processes including SOC and turbulence. Indeed, SOC models
exhibit threshold instabilities, bursty flow events and statistical features
consistent with the ``scale-free" dynamics such as power law power spectra.
It has been proposed by \cite{chang92b} that magnetospheric
dynamics are indeed in the critical state or near it. Alternatively,
\cite{consolini98} used the Castaing distribution -- the empirical model
derived in \cite{castaine} and based on a turbulent energy cascade -- to
obtain a two parameter functional form for the Probability Density Functions (PDF) of the AE fluctuations on various temporal scales. Turbulent descriptions
of magnetospheric measures also model observed statistical intermittency,
i.e., the presence of large deviations from the average value on different
scales \citep{consolini96,voros}. An increased probability of finding such
large deviations is manifested in the departure of the PDF from Gaussian
toward a leptokurtic distribution \citep{sornette}.

In this paper we will quantify both the intermittency and the self-similarity
of the AU, AL, AE and $\epsilon$ time series using the technique of finite
size scaling. This has the advantage of being model independent, and is also
directly related to both turbulence models such as that of Castaing \citep{castaine} and a Fokker-Planck description of the time series.
The method was used in \cite{hnat} where the mono-scaling of the solar wind
magnetic energy density fluctuations was reported. 
We will find that fluctuations in all four quantities are strongly suggestive
of turbulent processes and by quantifying this we can compare their properties
directly.

The AL, AU and AE indices data set investigated here comprises over $0.5$
million, $1$ minute averaged samples from January 1978 to January 1979
inclusive. The $\epsilon$ parameter defined in SI units as:
\begin{equation}
\epsilon=v \frac{B^2}{\mu_0}l_0^2 \sin^4(\Theta /2),
\label{eps}
\end{equation}
where $l_0 \approx 7R_E$ and $\Theta=\arctan(|B_y|/B_z)$
is an estimate of the fraction of the solar wind Poynting flux through the
dayside magnetosphere and was calculated from the WIND spacecraft key
parameter database \citep{lepping,ogilvie}. It comprises over $1$ million,
$46$ second averaged samples from January 1995 to December 1998 inclusive.
The data set includes intervals of both slow and fast speed streams.
The time series of indices and that of the $\epsilon$ parameter were obtained
in different time intervals and here we assume that the samples are long
enough to be statistically accurate.
\begin{figure}
\epsfsize=0.48\textwidth 
\centerline{
\leavevmode\epsffile{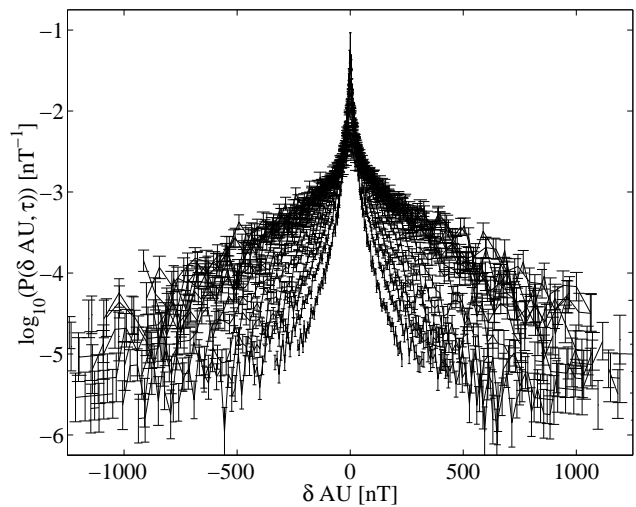}}
\caption{Unscaled PDFs of the AU index fluctuations. Time lag $\tau$
assumes values between $60$ seconds and about $36$ hrs. Standard deviation
of the PDF increases with $\tau$. Error bars on each bin within the PDF are
estimated assuming Gaussian statistics for the data within each bin.}
\label{fig1}
\end{figure}

\section{Scaling of the indices and $\epsilon$}
The statistical properties of complex systems can exhibit a degree 
of universality reflecting the lack of a characteristic scale in their
dynamics.
A connection between the statistical approach and the dynamical one is given
by a Fokker-Planck (F-P) equation \citep{kampen} which describes the dynamics
of the PDF and, in the most general form, can be written as:
\begin{equation}
\frac{\partial{P(x,t)}}{\partial{t}}=
\nabla (P(x,t) \gamma(x)) + \nabla^2 D(x) P(x,t), 
\label{eq1}
\end{equation}
where $P(x,t)$ is a PDF of some quantity $x$ that varies with time $t$,
$\gamma$ is the friction coefficient and $D(x)$ is a diffusion coefficient
which in this case can vary with $x$.
For certain choices of $D(x)$, a class of self-similar solutions of
(\ref{eq1}) satisfies a finite size scaling (in the usage of \cite{sornette},
pg. 85, henceforth ``scaling") relation given by:
\begin{equation}
P(x,t)=t^{-s} P_s(xt^{-s}).
\label{eq2}
\end{equation}
This scaling is a direct consequence of the fact that the F-P equation is
invariant under the transformation $x \rightarrow xt^{-s}$ and
$P \rightarrow Pt^s$.
If, for given experimental data, a set of PDFs can be constructed, on
different temporal scales $\tau$, that satisfy relation (\ref{eq2})
then a diffusion coefficient and corresponding F-P equation can be
found to represent the data. A simple example is the Brownian random walk
with $s=1/2$, $D(x)= $constant and Gaussian PDFs on all scales.
Alternatively one can treat the identification of the scaling exponent $s$
and, as we will see, the non-Gaussian nature of the rescaled PDFs ($P_s$)
as a method for quantifying the intermittent character of the time series.
Practically, obtaining the rescaled PDFs involves finding a rescaling index
$s$ directly from the integrated time series of the quantity $X$
\citep{hnat,sornette}.

Let $X(t)$ represent the time series of the studied signal, in our case
AU, AL, AE or the $\epsilon$ parameter. A set of time series
$\delta X(t,\tau)=X(t+\tau)-X(t)$ is obtained for each value of
non-overlapping time lag $\tau$. The PDF $P(\delta X,\tau)$ is then obtained
for each time series $\delta X(t,\tau)$. Figure \ref{fig1} shows these PDFs
for the $\delta AU$. A generic scaling approach is applied to these PDFs. 
Ideally, we use the peaks of the PDFs to obtain the scaling exponent $s$,
as the peaks are the most populated parts of the distributions.
In certain cases, however, the peaks may not be the optimal statistical
measure for obtaining the scaling index. For example, the $B_z$ component
in (\ref{eps}) as well as the AU and AL indices are measured with an
absolute accuracy of about $0.1$ nT. Such discreteness in the time series
and, in the case of the $\epsilon$ fluctuations, the large dynamical range
introduce large errors in the estimation of the peak values $P(0,\tau)$
and may not give a correct scaling. Since, if the PDFs rescale, we can
obtain the scaling exponent from any point on the curve in principle,
we also determine the scaling properties of the standard deviation
$\sigma(\tau)$ of each curve $P(\delta X, \tau)$ versus $\delta X(t,\tau)$.

\begin{figure}
\epsfsize=0.48\textwidth 
\centerline{
\leavevmode\epsffile{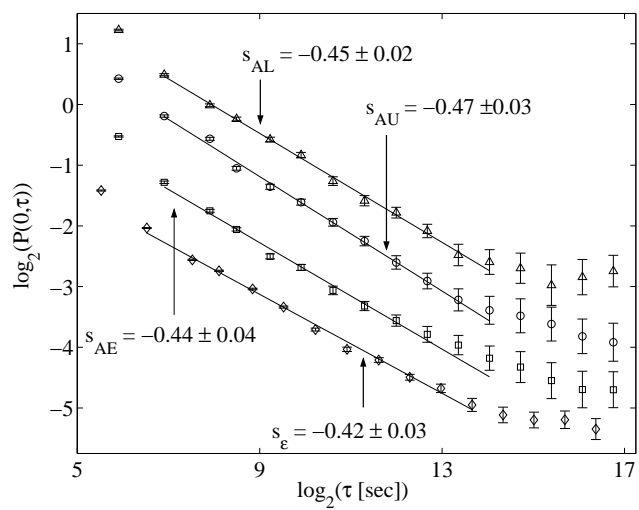}}
\caption{Scaling of the peaks of the PDFs for all quantities under.
investigation: $\diamond$ corresponds to $\epsilon$, $\circ$ AU index,
$\triangle$ AL index and $\Box$ the AE index. The plots have been offset
vertically for clarity. Error bars as in Figure 1.}
\label{fig2}
\end{figure}

Figure \ref{fig2} shows $P(0,\tau)$ plotted versus $\tau$ on log-log axes
for $\delta X= \delta \epsilon$, $\delta AE$, $\delta AU$ and $\delta AL$.
Straight lines on such a plot suggest that the rescaling (\ref{eq2}) holds
at least for the peaks of the distributions. On figure \ref{fig2}, lines
were fitted with $R^2$ goodness of fit for the range of $\tau$ between $4$
and $136$ minutes, omitting points corresponding to the first two temporal
scales as in these cases the sharp peaks of the PDFs can not be well
resolved. The lines suggest self-similarity persists up to intervals of
$\tau=97-136$ minutes. The slopes of these lines yield the exponents $s$ and
these are summarized in Table 1 along with the values obtained from analogous
plots of $\sigma(\tau)$ versus $\tau$ which show the same scale break.
We note that, for the $\epsilon$ parameter, the scaling index $s$ obtained
from the $P(0,\tau)$ is different from the Hurst exponent measured from the
$\sigma(\tau)$. This difference could be a result of the previously discussed
difficulties with the $\epsilon$ data. However, it does appear to be a feature
of some real time series (see \cite{gopikrishnan} for example).
Indeed, such a difference between index $s$ and $H_{\sigma}$ is predicted
in the case of fractional L\'{e}vy motion \citep{chechkin}.
We see that, for the $\epsilon$ as well as the AL and AU indices, there is a
range of $\tau$ up to $~4.5$ hours for which $P(0,\tau)$ plotted versus
$\tau$ is well described by a power law $\tau^{-s}$ with indices
$s=0.42 \pm 0.03$ for the $\epsilon$ and $s=0.45 \pm 0.02$ and
$s=0.47 \pm 0.03$ for the AL and AU indices, respectively.
Thus the break in scaling at $4-5$ hours in the AL and AU indices may have its
origin in the solar wind, although the physical reason for the break at this
timescale in epsilon is unclear.  The break in the AE index, however,
appears to occur at a smaller temporal scale of $~2$ hours, consistent with
the scaling break timescale found in the same index by other analysis methods
\citep{consolini98,takalo93}. This was interpreted by \citep{takalo93} as due
to the characteristic substorm duration. \cite{takalo98} also reported a
scaling break at the same $2$ hour timescale for AL, in contrast to the $4-5$
hour timescale found here. Indeed, one might have expected a substorm
timescale to cause the same scaling break in both the AE and AL indices,
because their substorm signatures are so similar in profile (e.g., Figure 2 of
\cite{caan}).  The resolution may lie in the difference between analysis of
differenced and undifferenced data \citep{price}.
\begin{table}[b]
\begin{center}
\begin{tabular}{|p{1.25cm}||p{2.0cm}|p{2.0cm}|p{1.2cm}|}
\hline
Quantity&$P(0,\tau)$ scaling index&$\sigma(\tau)$ scaling index&$\tau_{max}$\\
\hline 
$\epsilon$&$-0.42 \pm 0.03$&$0.33 \pm 0.04$&$~4.5$ hrs\\
\hline
AE-index&$-0.44 \pm 0.03$&$0.43 \pm 0.03$&$~2.1$ hrs\\
\hline
AU-index&$-0.47 \pm 0.03$&$0.47 \pm 0.02$&$~4.5$ hrs\\
\hline
AL-index&$-0.45 \pm 0.02$&$0.45 \pm 0.02$&$~4.5$ hrs\\
\hline
\end{tabular}
\caption{Scaling indices derived from $P(0,\tau)$ and $\sigma(\tau)$ power 
laws.}
\end{center}
\end{table}

\begin{figure}
\epsfsize=0.48\textwidth 
\centerline{
\leavevmode\epsffile{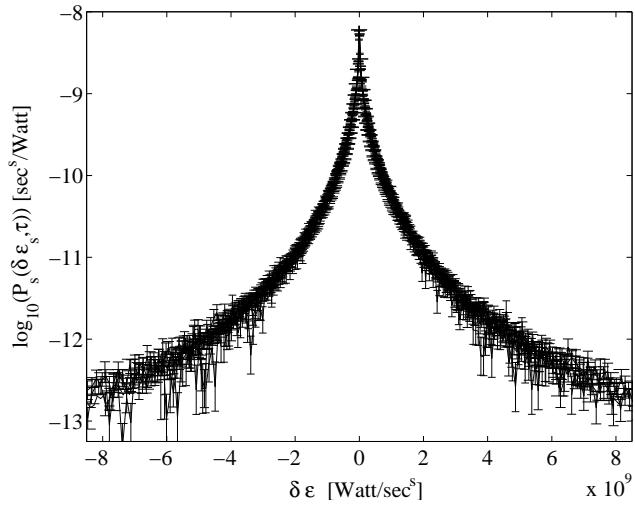}}
\caption{One parameter rescaling of the $\epsilon$ parameter fluctuations PDFs.
The curves shown correspond to $\tau$ between $46$ seconds and $~4.5$ hours.
Error bars as in Figure 1.}
\label{fig3}
\end{figure}

\begin{figure}
\epsfsize=0.48\textwidth 
\centerline{
\leavevmode\epsffile{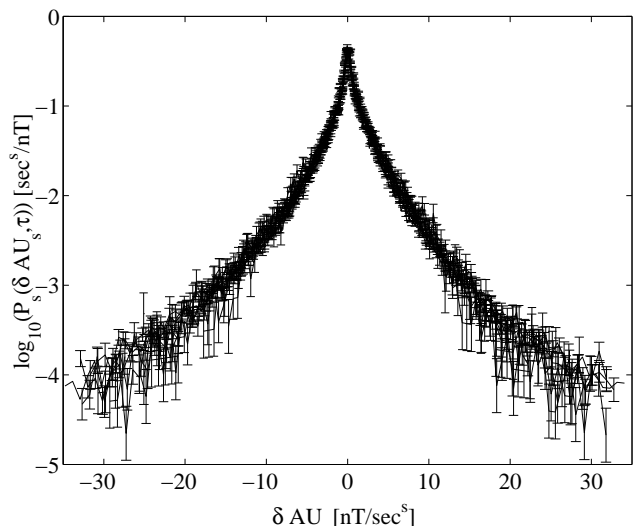}}
\caption{One parameter rescaling of the AU index fluctuation PDF.
The curves shown correspond to $\tau$ between $46$ seconds and $~4.5$ hours.
Error bars as in Figure 1.}
\label{fig4}
\end{figure}

Within this scaling range we now attempt to collapse each corresponding
unscaled PDF onto a single master curve using the scaling (\ref{eq2}).
If the initial assumption of the self-similar solutions is correct,
a single parameter rescaling, given by equation (\ref{eq2}) for a
mono-fractal process, would give a perfect collapse of PDFs on all scales.
Practically, an approximate collapse of PDFs is an indicator of
a dominant mono-fractal trend in the time series, i.e., this method may not
be sensitive enough to detect multi-fractality that could be present only
during short time intervals.
Figures \ref{fig3} and \ref{fig4} show the result of the one parameter
rescaling applied to the unscaled PDF of the $\delta \epsilon$ and the
$\delta AU$ index fluctuations, respectively, for the temporal scales up to
$\sim 4.5$ hours. We see that the rescaling procedure (\ref{eq2}) using the value of the exponent $s$ of the peaks $P(0,\tau)$ shown in Fig \ref{fig2},
gives good collapse of each curve onto a single common functional form for the
entire range of the data. These rescaled PDFs are leptokurtic rather than a
Gaussian and are thus strongly suggestive of an underlying turbulent process.

\begin{figure}
\epsfsize=0.48\textwidth 
\centerline{
\leavevmode\epsffile{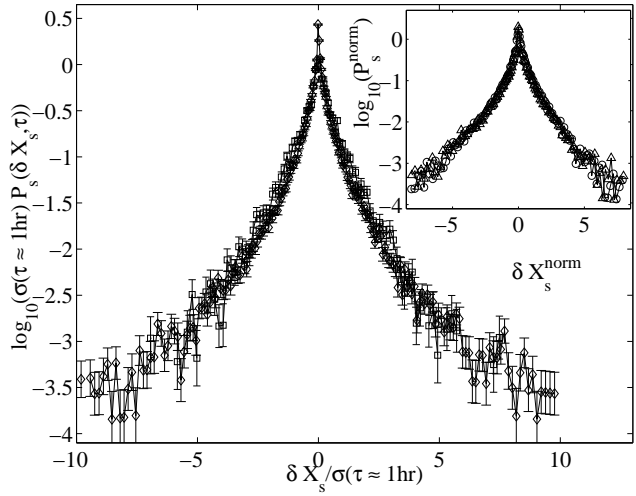}}
\caption{Direct comparison of the fluctuations PDFs for $\epsilon$
($\diamond$) and AE index ($\Box$). Insert shows overlaid PDFs of AU($\circ$)
and $-AL$($\triangle$) fluctuations. Error bars as in Figure 1.}
\label{fig5}
\end{figure}

The successful rescaling of the PDFs now allows us to perform a direct
comparison of the PDFs for all four quantities. Figure \ref{fig5} shows
these normalized PDFs $P_s(\delta X,\tau)$ for $\delta X=\delta \epsilon$,
$\delta AE$ and $\tau \approx 1$ hour overlaid on a single plot.
The $\delta X$ variable has been normalized to the rescaled
standard deviation $\sigma_s(\tau\approx 1hr)$ of $P_s$ in each case to
facilitate this comparison. We then find that AE and $\epsilon$ fluctuations
have indistinguishable $P_s$. The PDFs of $\delta AU$ and $\delta AL$ are
asymmetric such that $-\delta AL$ fits $\delta AU$ PDF closely (see insert
in the figure \ref{fig5}); when overlaid on the PDFs of the $\delta \epsilon$
and $\delta AE$ these are also indistinguishable within errors. This provides
strong evidence that the dominant contributions to the AE indices come from
the eastward and westward electrojets of the approximately symmetric DP2
current system that is driven directly by the solar wind \citep{freeman}.
The mono-scaling of the investigated PDFs, together with the finite value
of the samples' variance, indicates that a Fokker-Planck approach can be
used to study the dynamics of the unscaled PDFs within their temporal
scaling range.

\section{Summary}
In this paper we have applied the generic and model independent scaling method to study the scaling of fluctuations in the $\epsilon$
parameter and the global magnetospheric indices AU, AL and AE. 
The similar values of the scaling exponent and the leptokurtic nature of the
single PDF that, to within errors, describes fluctuations on time scales up to $\tau_{max}$ in all four quantities provide an important quantitative
constraint for models of the coupled solar wind-magnetosphere system.
One possibility is that, up to $\tau_{max} \sim4$ hours, fluctuations in
AU and AL are directly reflecting those  seen in the turbulent solar wind.
The data also suggest that AE index departs from this scaling on shorter
time scale of $\tau_{max} \sim2$ hours.
Importantly, identifying a close correspondence in the fluctuation PDF
of $\epsilon$, AE, AU and AL may simply indicate that fluctuations in the
indices are strongly coupled to dayside processes and are thus weak
indicators of the fluctuations in nightside energy output.
The leptokurtic nature of the PDFs is strongly suggestive of turbulent processes, and in the case of AU and AL, these may then be either just that of
the turbulent solar wind (and here $\epsilon$) or may be locally generated
turbulence which has an indistinguishable signature in its fluctuation PDF.
In this case our results quantify the nature of this turbulence. We note,
however, that certain classes of complex systems \citep{chang92a} are in
principle capable of ``passing through" input fluctuations into their output
without being directly driven in the present sense [Chang, private
communication, 2002].
Finally, the rescaling also indicates that a Fokker-Planck approach can be
used to study the evolution of the fluctuation PDF. This raises a possibility
of a new approach to understanding magnetospheric dynamics.
 
\section{Acknowledgment} 
SCC and BH acknowledge support from the PPARC and GR from the Leverhulme Trust.
We thank J. Greenhough and participants of the CEMD 2002 meeting for useful
discussions, and R. P. Lepping and K. Ogilvie for provision of data from the
NASA WIND spacecraft and the World Data Center C2, Kyoto for geomagnetic
indices. We thank anonymous referees for their comments.

\end{article}

\begin{thebibliography}{28}
\bibitem[{\it Angelopoulos et~al.}(1992)]{angelopoulos} Angelopoulos, V.
{\it et~al.}, Bursty bulk flows in the inner central plasma sheet, J. Geophys.
Res., {\bf 59}, 4027--4039, 1992.
\bibitem[{\it Caan et~al.}(1978)]{caan} Caan, M. N., R. L. McPherron, and
C. T. Russell, The statistical magnetic signature of magnetospheric substorms, 
{\em Planet. Space Sci. 26}, 269, 1978.
\bibitem[{\it Castaing et~al.}(1990)]{castaine} Castaing, B., Y. Gagne and E.J. Hopfinger, Velocity Probability Density Functions of High Reynolds Number
Turbulence, {\em Physica D, 46}, 177--200, 1990.
\bibitem[{\it Chang et~al.}(1992)]{chang92a} Chang, T. S., D. D. Vvedensky and
J. F. Nicoll, Differential renormalization-group generators for static and
dynamic critical phenomena, {\em Physics Reports, 217}, 279--362, 1992.
\bibitem[{\it Chang}(1992)]{chang92b} Chang, T., Low-dimensional Behavior and
Symmetry Breaking of Stochastic Systems Near Criticality: Can these Effects
be Observed in Space and in the Laboratory?, {\em IEEE Trans. Plasma Sci., 20},
691, 1992.
\bibitem[{\it Chapman and Watkins}(2001)]{chapman01} Chapman, S. C., and N. W.
Watkins, Avalanching and Self Organised Criticality: a paradigm for
magnetospheric dynamics?, {\em Space Sci. Rev., 95}, 293--307, 2001.
\bibitem[{\it Chechkin and Gonchar}(2000)]{chechkin} Chechkin, A. V., and V. Yu.
Gonchar, A model for persistent Levy motion, {\em Physica A, 277}, 312--326,
2000.
\bibitem[{\it Consolini et~al.}(1996)]{consolini96} Consolini, G., M. F.
Marcucci, M. Candidi, Multifractal structure of auroral electrojet index data,
{\em Phys. Rev. Lett., 76}, 4082--4085, 1996.
\bibitem[{\it Consolini and De Michelis}(1998)]{consolini98} Consolini, G.,
and P. De Michelis, Non-Gaussian distribution function of AE-index
fluctuations: Evidence for time intermittency, {\em Geophys. Res. Lett., 25},
4087--4090, 1998.
\bibitem[{\it Davis and Sugiura}(1966)]{davis} Davis, T. N., and M. Sugiura,
Auroral electrojet activity index {\it AE} and its universal time variations,
{\em J. Geophys. Res., 71}, 785--801, 1966.
\bibitem[{\it Freeman et~al.}(2000)]{freeman} Freeman, M. P., N.~W. Watkins and D.J. Riley, Evidence for a solar wind origin of the power law burst
lifetime distribution of the AE indices, 
\grl {\em 27}, 1087--1090, 2000.
\bibitem[{\it Gopikrishnan et~al}(1999)]{gopikrishnan} Gopikrishnan, P.,
V. Plerou, L. A. Nunes Amaral, M. Meyer and H. E. Stanley,
Scaling of the distribution of fluctuations of financial market indices,
{\em Phys. Rev. E, 60}, 5305--5316, 1999.
\bibitem[{\it Hnat et~al.}(2002)]{hnat} Hnat, B., S. C. Chapman, G. Rowlands,
N. W. Watkins, W. M. Farrell, Finite size scaling in the solar wind magnetic
field energy density as seen by WIND, {\em Geophys. Res. Lett., 29(10)},
10.1029/2001GL014587, 2002.
\bibitem[{\it Klimas et~al.}(1996)]{klimas} Klimas, A. J., D. Vassiliadis,
D. N. Baker and D. A. Roberts, The organized nonlinear dynamics of the
magnetosphere, {\em J. Geophys. Res., 101}, 13089--13113, 1996.
\bibitem[{\it Lepping et~al.}(1995)]{lepping} Lepping, R. P., M. Acuna, L.
Burlaga, W. Farrell, J. Slavin, K. Schatten, F. Mariani, N. Ness, F. Neubauer,
Y. C. Whang, J. Byrnes, R. Kennon, P. Panetta, J. Scheifele, and E. Worley,
The WIND Magnetic Field Investigation, {\em Space Sci. Rev., 71}, 207, 1995.
\bibitem[{\it Lui et~al.}(2000)]{lui} Lui, A. T. Y., {\it et~al.},
Is the Dynamic Magnetosphere an Avalanching System?,
{\em Geophys. Res. Lett., 27}, 911--914, 2000.
\bibitem[{\it Ogilvie et~al.}(1995)]{ogilvie} Ogilvie, K. W., D. J. Chornay,
R. J. Fritzenreiter, F. Hunsaker, J. Keller, J. Lobell, G. Miller, J. D.
Scudder, E. C. Sittler, R. B. Torbert, D. Bodet, G. Needell, A. J. Lazarus,
J. T. Steinberg, J. H. Tappan, SWE, a comprehensive plasma instrument for the wind spacecraft, {\em Space Sci. Rev., 71}, 55--77, 1995.
\bibitem[{\it Perreault and Akasofu}(1978)]{akasofu} Perreault, P., and S.-I.
Akasofu, A study of geomagnetic storms, {\em Geophys. J. R. Astr. Soc, 54},
547--573, 1978.
\bibitem[{\it Price and Newman}(2001)]{price} Price, C. P., and D. E. Newman,
Using the R/S statistic to analyze AE data,
{\em J. Atmos. Sol.-Terr. Phys., 63}, 1387--1397,  2001.
\bibitem[{\it Sitnov et~al.}(2000)]{sitnov} Sitnov, M. I., A. S. Sharma, K. 
Papadopoulos, D. Vassiliadis, J. A. Valdivia, A. J. Klimas, D. N. Baker,
Phase transition-like behavior of the magnetosphere during substorms,
\jgr {\em 105}, 12955--12974, 2000.
\bibitem[{\it Sornette}(2000)]{sornette} Sornette, D.,
{\em Critical Phenomena in Natural Sciences; Chaos, Fractals, Selforganization
and Disorder: Concepts and Tools}, Springer-Verlag, Berlin, 2000.
\bibitem[{\it Takalo et~al.}(1993)]{takalo93} Takalo, J., J. Timonen., and
H. Koskinen, Correlation dimension and affinity of AE data and bicolored
noise, {\em Geophys. Res. Lett., 20}, 1527--1530, 1993.
\bibitem[{\it Takalo and Timonen}(1998)]{takalo98} Takalo J., and J. Timonen,
Comparison of the dynamics of the AU and PC indices,
{\em Geophys. Res. Lett., 25}, 2101-2104, 1998.
\bibitem[{\it Tsurutani et~al.}(1990)]{tsurutani} Tsurutani, B. T., M. Sugiura,
T. Iyemori, B. E. Goldstein, W. D. Gonzalez, S. I. Akasofu, E. J. Smith,
The nonlinear response of AE to the IMF $B_s$ driver: A spectral break at $5$
hours, {\em Geophys. Res. Lett., 17}, 279--282, 1990.
\bibitem[{\it Uritsky et~al.}(2001)]{uritsky} Uritsky,  V. M., A. J. Klimas
and D. Vassiliadis, Comparative study of dynamical critical scaling in the
auroral electrojet index versus solar wind fluctuations,
{\em Geophys. Res. Lett., 28}, 3809--3812, 2001.
\bibitem[{\it van Kampen}(1992)]{kampen} van Kampen, N.G., {\em Stochastic Processes in Physics and Chemistry}, North-Holland, Amsterdam, 1992
\bibitem[{\it Vassiliadis et~al.}(2000)]{vassiliadis} Vassiliadis, D., A. J.
Klimas, J. A. Valdivia, and D. N. Baker, The Nonlinear Dynamics of Space
Weather, {\em Adv. Space. Res., 26}, 197--207, 2000.
\bibitem[{\it V\"{o}r\"{o}s et~al.}(1998)]{voros} V\"{o}r\"{o}s, Z.,
P. Kov\'{a}cs, \'{A}. Juh\'{a}sz, A. K\"{o}rmendi and A. W. Green, Scaling laws
from geomagnetic time series, {\em Geophys. Res. Lett., 25}, 2621-2624, 1998.
\end{thebibliography}
\end{document}